\documentclass[10pt]{article}

\usepackage{amsmath}
\usepackage{amssymb}

\newcommand{\hl}[1]{#1}
\usepackage{graphicx}

\usepackage{cite}

\usepackage{color} 

\usepackage[labelfont=bf,labelsep=period,justification=raggedright]{caption}

\makeatletter
\renewcommand{\@biblabel}[1]{\quad#1.}
\makeatother

\date{}

\usepackage{verbatim}
\usepackage{amsmath}
\usepackage{amssymb}
\usepackage{tabularx}
\usepackage{graphicx}

\usepackage{cite}

\usepackage{color}

\usepackage{setspace} 
\usepackage{ifthen}

\usepackage[boxed,noend]{algorithm2e}

\usepackage{hyperref}

\makeatletter
\renewcommand{\@biblabel}[1]{\quad#1.}
\makeatother

\date{}

\newcommand\titlestring{Developing and applying heterogeneous phylogenetic models with XRate}
\newcommand\shorttitlestring{Phylogenetic modeling with XRate}

\usepackage{setspace}
\usepackage{array}

\newcommand{\supptext}[1]{Text S#1}

\newcommand{\secref}[1]{Section~\ref{sec.#1}}
\newcommand{\seclabel}[1]{\label{sec.#1}}

\newcommand{\figref}[1]{Figure~\ref{Figures.#1}}
\newcommand{\figlabel}[1]{\label{Figures.#1}}

\newcommand\xrate{XRate}
\newcommand\xrateformat{\url{http://biowiki.org/XrateFormat}}
\newcommand\xrateMacros{\url{http://biowiki.org/XrateMacros}}

\newcommand\paperBiowikiPage{ \url{http://biowiki.org/XratePaper2011}}

\begin{document}

\begin{flushleft}
{\Large
\titlestring
}
\\
Oscar Westesson$^{1}$, 
Ian Holmes$^{2,\ast}$
\\
Department of Bioengineering, University of California, Berkeley, CA, USA
\\
\textbf{1}  breadbaron@berkeley.edu\\
\textbf{2}  ihh@berkeley.edu\\
$\ast$ Corresponding author 
\end{flushleft}

\section*{Abstract}
Modeling sequence evolution on phylogenetic trees is a useful technique in computational biology.
Especially powerful are models which take account of the heterogeneous nature of sequence evolution
according to the ``grammar'' of the encoded gene features.
However, beyond a modest level of model complexity, manual coding of models becomes prohibitively labor-intensive.  \\
We demonstrate, via a set of case studies, the new built-in model-prototyping capabilities of
 \xrate\ (macros and Scheme extensions). These features allow rapid implementation of phylogenetic models
 which would have previously been far more labor-intensive.  
\xrate's new capabilities for
 lineage-specific models, ancestral sequence reconstruction, and improved annotation output are also discussed. \\
 \xrate's flexible model-specification capabilities and computational efficiency
make it 
well-suited to developing and prototyping phylogenetic grammar models.
\xrate\ \hl{is available as part of the DART software package:} \url{http://biowiki.org/DART} .


\section*{Introduction}
Phylogenetics, the modeling of  evolution on trees, is an extremely powerful tool in computational biology.
  The better we can model a system, the more can learn from it, and vice-versa.
  Especially attractive, given the plethora of available sequence data, is modeling sequence evolution at the molecular level.
 Models describing the evolution of a single nucleotide began simply (e.g. JC69 \cite{JukesCantor69}), later evolving to capture such biological features as transition/transversion bias (e.g. K80 \cite{Kimura80}) and unequal base frequencies (e.g. HKY85 \cite{HasegawaEtAl85}).
  Felsenstein's ``pruning'' algorithm allows combining these models with phylogenetic trees
to compute the likelihood of multiple sequences
 \cite{Felsenstein81}. 

As powerful as phylogenetic models are for explaining the evolutionary depth of a sequence alignment, they are even more powerful when combined with a model for the feature structure: \hl{the partition of the alignment into regions, each evolving under a particular model. }
The phylogenetic grammar, or ``phylo-grammar'', is one such class of models.
Combining hidden Markov models (and, more generally, stochastic grammars) and phylogenetic substitution models provides computational modelers with a rich set of comparative tools to analyze multiple sequence alignments (MSAs): gene prediction, homology detection, finding structured RNA, and detecting changes in selective pressure have all been approached with this general framework \cite{MeyerDurbin04,Eddy98,KnudsenHein99,GarberEtAl2009}.  
\hl{Readers unfamiliar with phylo-grammars may benefit from relevant descriptions 
and links available here:} \url{http://biowiki.org/PhyloGrammars} or the original
paper describing \xrate \cite{KlostermanEtAl2006}.
\hl{Also, a collection of animations depicting various evolutionary models
 at work (generating multiple alignments or evolving sequences) has been 
compiled here:} \url{http://biowiki.org/PhyloFilm} .

While the mathematics of sequence modeling is straightforward, manual implementation can quickly become the limiting factor in iterative development of a
 computational pipeline. 
To streamline this step, general modeling platforms have been developed. For instance, Exonerate allows users to specify a wide variety of common substitution and gap models when aligning pairs of sequences \cite{SlaterBirney2005}.  
Dynamite uses a specification file to generate code for dynamic programming routines \cite{BirneyDurbin97}. HMMoC is a similar model compiler sufficiently general to work with arbitrary HMMs \cite{Lunter2007}.
  The BEAST program allows users to choose from a wide range of phylogenetic substitution models while also sampling over trees \cite{DrummondRambaut2007}.
  The first three of these are non-phylogenetic, only able to model related pairs of sequences.
  Dynamite and HMMoC are unique in that they allow definition of arbitrary models via specification files, whereas users of BEAST and Exonerate are limited to the range of models which have been hard-coded in the respective programs.  

Defining models' structure manually can be limiting as models grow in size and/or complexity.  
For instance, a Nielsen-Yang model incorporating both selection and transition/transversion bias has nearly 4000 entries - far too many for a user to manually specify \cite{YangEtAl2000}.
  Such a large matrix requires specific model-generating code to be written and integrated with the program in use - not always possible or practical for the user depending on the program's implementation. 

\xrate\ is a phylogenetic modeling program that implements the key parameterization and inference algorithms given two ingredients: a user-specified phylo-grammar, and a multiple sequence alignment.
(A phylogeny can optionally be specified by the user, or it can be inferred by the program.)
\xrate's models describe the parametric structure of substitution rate matrices, along with grammatical rules governing which rate matrices can account for which alignment columns.
\hl{This essentially amounts to partitioning the alignment (e.g. marking up exon boundaries and reading frames)
 and factoring in the transitions between the different types of region.}

  Parameter estimation and decoding (alignment annotation) algorithms are built in, allowing fast model prototyping and fitting.  \hl{Model training (estimating the rate and probability
parameters of the grammar) is done via a form of the Expectation Maximization (EM) algorithm,
 described in more detail in the original} \xrate\ \hl{paper} \cite{KlostermanEtAl2006}.
  Most recently, \xrate\ allows programmatic model construction via its macros and Scheme extensions.
 \xrate's built-in macro language allows large, repetitive grammars to be compactly represented, and also enables the model structure to depend on aspects of the data, such as the tree or alignment.
  Scheme extensions take this even further, interfacing \xrate\ to a full-featured functional scripting language, allowing complex \xrate-oriented workflows to be written as Scheme programs. 

In this paper we demonstrate \xrate's new model-specification tools via a set of progressively more complex examples, concluding with XDecoder, a phylo-grammar modeling RNA secondary structure overlapping protein-coding regions. 
We also describe additional improvements to \xrate\ since its initial publication, namely ancestral sequence reconstruction, GFF/WIG output, and hybrid substitution models.  
Finally, we show how \xrate's features are exposed as function extensions in a dialect of the Scheme programming language,
typifying a Functional Programming (FP) style of model development and inference for phylogenetic sequence analysis.
Terminology relevant to modeling with \xrate\ are defined in detail in Appendix \secref{glossary}.
\hl{We also provide an online tutorial for making nontrivial modifications to existing grammars, going step-by-step from a Jukes-Cantor model to an autocorrelated Gamma-distributed rates phylo-HMM:} \url{http://biowiki.org/XrateTutorial}. 

\section*{Results and Discussion}

\subsection*{The \xrate\ generative model}
A phylo-grammar generates an  alignment in two steps: nonterminal transformations and token evolution. 
The sequence of nonterminal transformations comprises the ``grammar'' portion of a phylo-grammar, and the ``phylo'' portion refers to the evolution of tokens along a phylogeny.
First, transformation rules are repeatedly applied, beginning with the {\tt START} nonterminal, until only a series of pseudoterminals remains.
From each group of pseudoterminals (a group may be a single column, two ``paired'' columns in an RNA structure, or a codon triplet of columns), a tuple of tokens is sampled from the initial distribution of the chain corresponding to the pseudoterminal.
  These tokens then evolve down the phylogenetic tree according to the mutation rules of the chain, resulting in the observed alignment columns. 

If the nonterminal transformations contain no bifurcations and all
 emissions occur on the same side of the nonterminal, the grammar is a phylogenetic hidden Markov model (phylo-HMM), a special subclass of phylo-grammars.  
Otherwise, it is a phylogenetic stochastic context-free grammar (phylo-SCFG), the most general class of models implemented by \xrate. 
  This distinction, along with other related technical terms, are described in greater detail in Appendix \secref{glossary}, the Glossary of \xrate\ model terminology.

\hl{The generality of } \xrate\ \hl{requires a slight tradeoff against speed.  
Since the low-level code implementing core operations
is shared among the set of possible  models,} \xrate\ \hl{will generally be slower than
programs with source code optimized for a narrower range of models. 
Computing the Felsenstein likelihood under the HKY85} \cite{HasegawaEtAl85} \hl{model of a 5-taxon, 1Mb alignment,} \xrate\ \hl{required 1.25 minutes of CPU time and 116MB  RAM, while 
PAML required 9 seconds of CPU time and 19MB RAM for the same operation. 
Running PFOLD} \cite{KnudsenHein2003} \hl{on a 5-taxon, 1KB alignment required 11 seconds and 164MB RAM, and running} \xrate\  \hl{on the same alignment with a comparable grammar 
required 25 seconds and 62MB RAM. 
All programs were run with default settings on a 3.4 GHz Intel i7 processor. 
Model-fitting also takes longer with \xrate:
a previous work found that \xrate's parameter estimation routines were approximately 130 times
slower than those in PAML} \cite{HegerEtAl2009}.  

\hl{In an attempt to improve } \xrate's \hl{performance, we tried using Beagle,
a library that provides CPU and accelerated parallel GPU implementations of Felsenstein's algorithm
along with related matrix operations} \cite{AyresEtAl2011}.  
\hl{We have, however, been so far unable to generate significant performance gains by this method.}

\hl{Despite these caveats,} \xrate\ \hl{ has proved to be fast enough for genome-scale
applications, such as a screen of {\em Drosophila} whole-genome alignments} \cite{BradleyUzilovEtAl2009}.
\hl{Furthermore, it implements a significantly broader range of models than the above-cited tools.}



\subsection*{\xrate\ inputs, outputs and operations}
\hl{The formulation of the }\xrate\ \hl{model presented in the previous section
is generative: that is, it describes the generation of data on a tree.
In practice, the main reason for doing this is to generate simulation data for benchmarking purposes.
This is possible using the tool {\tt simgram}} \cite{VaradarajanEtAl2008},
\hl{which is provided with }\xrate\ \hl{as part of the DART package.}

\hl{Most common use cases for generative models involve not simulation, but inference:
that is, reconstructing aspects of the generative process (sequence of nonterminal transformations,  token mutations, or grammar parameters) given observed sequence data (in the form of a multiple sequence alignment). }
Using a phylo-grammar, a set of aligned sequences, and a phylogeny relating these sequences (optionally inferred by \xrate), \xrate\ implements the relevant  parameterization and inference algorithms, allowing researchers to analyze sequence data without having to implement their own models.  

Sequences are read and written in Stockholm format \cite{StkFormat} (converters to and from common formats are included with DART). \hl{This format allows for the option of  embedding a tree in Newick format} \cite{NewickFormat} (via the \verb|#=GF NH| tag) and annotations in GFF format \cite{GFF98}.  
\hl{By construction, Newick format necessarily specifies a rooted tree, rather than an unrooted one.
However, the root placement is only relevant for time-irreversible models; 
when using time-reversible models, the placement of the root is arbitrary and can safely be ignored.}
Given these input ingredients, a call to \xrate\ proceeds in the following order 
\hl{(more detail is provided at} \url{http://biowiki.org/XRATE} and \url{http://biowiki.org/XrateFormat} ):
\begin{enumerate}
\item The Stockholm file and grammar alphabet are parsed (as macros may depend on these).
\item Any grammar macros are expanded, \hl{followed by Scheme functions.}
\item If requested, or a tree was not provided in the input data, one is estimated using neighbor-joining \cite{SaitouNei87}. \hl{As noted above, this is a rooted tree, but the root placement is arbitrary if a time-reversible model is used.}
\item Grammar parameters are estimated (if requested).
\item Alignment is annotated (if requested).
\item Ancestral sequences are reconstructed (if requested).
\end{enumerate}

After the analysis is complete, the alignment (along with an embedded tree) is printed to the output stream along with ancestral sequences (if requested) as well as any \verb|#=GC| and \verb|#=GR| column annotations.
GFF and WIG annotations are  sent to standard output by default, but these can be directed to separate files by way of the {\tt -gff} and {\tt -wig} options, respectively. 


\subsection*{The \xrate\ format macro language for phylo-grammar specification: case studies}

The following sections describe case studies of repetitively-structured models which motivate the need for grammar-generating code.
Historically, we have attempted several solutions to the case studies described.
We first briefly review the factors that influenced our eventual choice of Scheme as a macro language.

\xrate\ was preceded by Searls' Prolog-based automata \cite{Searls95} and Birney's Dynamite parser-generator \cite{BirneyDurbin97},
and roughly contemporaneous with Slater's Exonerate \cite{SlaterBirney2005} and Lunter's HMMoC \cite{Lunter2007}.
In early versions of \xrate\ (circa 2004), and in Exonerate, the only way for the user to specify their own phylo-grammar models
was to write C/C++ code that would compile directly against the program's internal libraries.
This kind of compilation step significantly slows model prototyping, and impedes re-use of model parameters.

Current versions of \xrate, along with Dynamite and HMMoC, understand a machine-readable grammar format.
In the case of \xrate, this format is based on Lisp S-expressions.
In such formats (as the case studies illustrate) the need arises for code that generates repetitively-structured grammar files.
It is often convenient, and sometimes sufficient, to write such grammar-generating code in an external language:
for example, we have written Perl, Python and C++ libraries to generate
\xrate\ grammar files \cite{KlostermanEtAl2006,HegerEtAl2009}.
However, this approach still has the disadvantage (from a programmer's or model developer's perspective) that
(a) code to generate real grammars tends to require an ungainly mix of grammar-related S-expression constants embedded in Perl/Python/C++ code,
and
(b) the requirement for an explicit model-generation step can delay prototyping and evaluation of new phylo-grammar models.

\xrate's macro language  provides an alternate way to generate repetitive models within \xrate,
without having to resort to external code-generating scripts.  This allows the  model-specifying code to remain compact, readable, and easy to edit.
As we report in this manuscript, the \xrate\ grammar format now also natively includes a Scheme-based
scripting language that can be embedded directly within grammar files,
whose syntax blends seamlessly with the S-expression format used by \xrate\ and whose functional nature fits \xrate's problem domain.
We provide here examples of common phylogenetic models which make use of various macro features, and refer the reader to the online documentation for a complete introduction to \xrate's macro features: \xrateMacros.  All of the code snippets presented here are available as minimal  complete grammars in \supptext{1}.  The full, trained grammars corresponding to those presented here are available as part of DART.  This correspondence is described here: \paperBiowikiPage

\subsubsection*{A repetitively-structured HMM specified using simple macros}
\seclabel{modelRedundancy}

Probabilistic models for the evolution of biological sequences tend to contain repetitive structure.
Sometimes, this structure arises as a reflection of symmetries in the phylo-grammar; other times, it arises due to structure in the data, such as the tree or the alignment. 
While small repetitive models can be written manually, 
developing richer evolutionary models and grammars often demands  writing code to model the underlying structure.

\paragraph{Markov chain symmetry} The most familiar source of repetition derives from the substitution model's structure: different substitutions share parameters based on prior knowledge or biological intuition.  Perhaps most repetitive is the Jukes-Cantor model for DNA. The matrix entries $Q_{ij}$ denote the rate of substitution from $i$ to $j$:

\[
Q^{JC} = \left\{
\begin{array}{ccccc}
 & \mbox{{\bf A}} & \mbox{{\bf C}} & \mbox{{\bf G}} & \mbox{{\bf T}}\\
\mbox{{\bf A}} & * & u & u & u \\
\mbox{{\bf C}} & u & * & u & u \\
\mbox{{\bf G}} & u & u & * & u \\
\mbox{{\bf T}} & u & u & u & * \\
\end{array}
\right\}
\]

Here $u$ is an arbitrary positive rate parameter.
The $*$ character denotes the negative sum of the remaining row entries (here equal to $-3u$ in every case).
The parameter $u$ is typically set to $1/3$ in order that the stochastic process performs, on average,
one substitution event per unit of time.

This matrix can be specified in \xrate\ with two nested loops over alphabet tokens.
Each loop over alphabet tokens has the form \verb|(&foreach-token X expression...)|
where \verb|expression...| is a construct to be expanded for each alphabet token \verb|X|.
\hl{Here,} \verb|expression| \hl{sets  the substitution rate} between each pair of source and destination tokens
(except for the case when the source and destination tokens are identical,
for which case we simply generate an empty list, \verb|()|, which will be ignored by the \xrate\ grammar parser).
We do not explicitly need to write the negative values of the on-diagonal matrix elements
(labeled $*$ in the above description of the matrix); \xrate\ will figure these out for itself.
To check whether source and destination tokens are equal in the loop, we use a conditional \verb|&if| statement,
which has the form \verb|(&if (condition) (expansion-if-true) (expansion-if-false))|.
The \verb|condition| is implemented using the \verb|&eq| macro, which tests if its two arguments are equal.
Putting all these together, the nested loops look like this:

\begin{verbatim}
(&foreach-token tok1
 (&foreach-token tok2
  (&if (&eq tok1 tok2) 
   ()  ;; If tok1==tok2, expand to an empty list (ignored by parser)
   (mutate (from (tok1)) (to (tok2)) (rate u)))))
\end{verbatim}


While this illustrates \xrate's looping and conditional capabilities, such a simple model would almost be easier to code by hand.
  For a slightly more complex application, we turn to the model of Pupko {\em et al} in their 2008 work.
  In their RASER program the authors used a chain augmented with a latent variable indicating ``slow'' or ``fast'' substitution.
  Reconstructing ancestral sequences on an HIV phylogeny allowed them to infer locations of transitions between slow and fast modes - indicating a possible gain or loss of selective pressure \cite{PennEtAl2008}.
   The chain shown below, $Q^{RASER}$, shows a simplified version of their model: substitutions {\em within} rate classes occur according to a JC69 model scaled by rate parameters $s$ and $f$ (slow and fast, respectively), and transitions {\em between} rate classes occur with rates $r_{sf}$  and $r_{fs}$ (slow $\rightarrow$ fast and fast $\rightarrow$ slow, respectively).

\[
Q^{RASER} = \left\{
\begin{array}{ccccc|cccc}
 & \mbox{{\bf A$_s$}} & \mbox{{\bf C$_s$}} & \mbox{{\bf G$_s$}} & \mbox{{\bf T$_s$}} & \mbox{{\bf A$_f$}} & \mbox{{\bf C$_f$}} & \mbox{{\bf G$_f$}} & \mbox{{\bf T$_f$}}\\
\mbox{{\bf A$_s$}} & * & us & us  & us & r_{sf} & 0 & 0 & 0 \\
\mbox{{\bf C$_s$}} &  us & * & us  & us & 0 &r_{sf} & 0 & 0 \\
\mbox{{\bf G$_s$}} & us & us & *  & us & 0 & 0 & r_{sf} & 0 \\
\mbox{{\bf T$_s$}} & us & us & us  & * & 0 & 0 & 0 &r_{sf}  \\
\hline
\mbox{{\bf A$_f$}} & r_{fs} & 0 & 0 & 0 & * & uf & uf  & uf \\
\mbox{{\bf C$_f$}} & 0 &r_{fs} & 0 & 0 &  uf & * & uf  & uf \\
\mbox{{\bf G$_f$}} & 0 & 0 & r_{fs} & 0 & uf & uf & *  & uf \\
\mbox{{\bf T$_f$}} & 0 & 0 & 0 &r_{fs} & uf & uf & uf  & * \\
\end{array}
\right\}
\]

While this chain contains  four times as many rates as the basic JC69 model, there are only five parameters: $u,s,f,r_{sf}, r_{fs}$ since the model contains repetition via its symmetry.
  While manual implementation is possible, the model can be expressed in just a few lines of \xrate\ macro code.
  Further, additional ``modes'' of substitution (corresponding to additional quadrants in the matrix above) can be added by editing the first two lines of the following code.  

\xrate\ represents latent variable chains as tuples of the form {\tt (state class)}, where {\tt state} is a particular state of the Markov chain and {\tt class} is the value of a hidden variable.  
In this case, standard DNA characters are augmented with a latent variable indicating substitution rate class: ${\bf A}_f$ indicates an {\bf A} which evolves ``fast.''  
The following syntax is used to declare a latent variable chain (in this case, this variable may take values {\tt s} or {\tt f}), with the {\tt row} tag  specifying {\tt CLASS} as the
Stockholm \verb|#=GR| identifier for per-sequence, per-column annotations:

{\tt  (hidden-class (row CLASS) (label (s f)))}

Combining loops, conditionals, hidden classes, and the \verb|(&cat LIST)| function (which concatenates the elements of {\tt LIST}), we get the following \xrate\ code for the RASER chain:

\begin{verbatim}
(rate (s 0.1) (f 2.0) (r_sf 0.01) (r_fs 0.01) (u 1.0))
(chain  
 (hidden-class (row CLASS) (label (s f)))
 (terminal RASER)
 (&foreach class1 (s f)
  (&foreach class2 (s f)
   (&foreach-token tok1
    (&foreach-token tok2
     (&if (&eq class1 class2)
      (&if (&eq tok1 tok2)
       () ;; if class1==class2 && tok1==tok2, expand to empty list (will be ignored)
       ;; The following line handles the case (class1==class2 && tok1!=tok2)
       (mutate (from (tok1 class1)) (to (tok2 class2)) (rate u class1)))
      (&if (&eq tok1 tok2)
       ;; The following line handles the case (class1!=class2 && tok1==tok2)
       (mutate (from (tok1 class1)) (to (tok2 class2)) (rate (&cat r_ class1 class2)))
       ())))))))  ;; if class1!=class2 && tok1!=tok2, expand to empty list (ignored)
\end{verbatim}

\paragraph{Phylo-HMM-induced repetition}

The previous examples both involved specifying the Markov chain component of a phylo-grammar. Coupled with a trivial top-level grammar
 (a {\tt START} state and an {\tt EMIT} state which emits the chain via the {\tt EMIT*} pseudoterminal), 
 these models describe an alignment where each column's characters evolve according to the same substitution model. 
A common extension to this is using sequences of hidden states which generate alignment columns according to different substitution models. 
These ``phylo-grammars'' (which can include phylo-SCFGs and the more restricted phylo-HMMs)
allow modelers to describe and/or detect alignment regions exhibiting different evolutionary patterns.
  Phylo-HMMs model left-to-right correlations between alignment columns, and phylo-SCFGs are capable of modeling nested correlations (such as ``paired'' columns  in an RNA secondary structure).  
\hl{Readers unfamiliar with phylo-grammars may benefit from relevant descriptions 
and links available here:} \url{http://biowiki.org/PhyloGrammars}, \hl{animations available here:} \url{http://biowiki.org/PhyloFilm}, \hl{and the original paper describing} \xrate \cite{KlostermanEtAl2006}.

We outline here a phylo-HMM that is simple to describe, but would take a substantial amount of code to implement without \xrate's macro language. 
The model is based on PhastCons, a program by Siepel {\em et al} which uses an HMM whose three states (or, in \xrate\ terminology, nonterminals)
use substitution models differing only by rate multipliers \cite{SiepelEtAl2005}.
This model, depicted schematically in  \figref{phastconsDiagram}, can be used to detect alignment regions evolving at different rates.
\hl{If the rates of each hidden state correspond to quantiles of the Gamma distribution, then 
summing over hidden states of this model is equivalent to the commonly-used Gamma model of
rate heterogeneity.  
We provide this grammar in} \supptext{1}\hl{, which is essentially identical to the PhastCons 
grammar with $n$ states except for its invocation of a Scheme function returning the $n$ 
Gamma-derived rates for a given shape parameter. }
  We can define such a model in \xrate\ easily due to the symmetric structure:
all three nonterminals have similar underlying substitution models
 (varying only by a multiplier)
and also similar probabilities of making transitions to other nonterminals via grammar transformation rules.



The grammar will have nonterminals named ``1'', ``2''...up to {\tt numNonTerms}, each one associated with a rate parameter ({\tt r\_1, r\_2...}) and substitution chain ({\tt chain\_1, chain\_2...}).  
To express this grammar in \xrate\ macro code, we'll need to declare each of these nonterminals, the production rules which govern transitions between them, rate parameters, and the nonterminal-associated substitution chains.  
(For a fully-functional grammar, an alphabet is also needed; these are omitted in code snippets included in the  main text, but the corresponding grammars in \supptext{1} contain alphabets.)


First, define how many nonterminals the model will have:
adding more nonterminals to the model later on can be done simply by adjusting this variable.
We define a \verb|SEED| value to initialize the rate parameters \hl{(this is not a random  
number seed, but rather an initial guess at the parameter value necessary for the EM algorithm
to begin)},
which is done inside a \verb|foreach-integer| loop using the \verb|numNonterms| variable.  
The  {\tt (foreach-integer X (1 K) expression)} expands {\tt expression} for all values of {\tt X} from 1 to {\tt K}.  
In this case, we define a rate parameter for each of our nonterminals 1..{\tt K}.

\begin{verbatim}
(&define numNonterms 3)
(&define SEED 0.001)
(&foreach-integer nonterminal (1 numNonterms)
    (rate ((&cat r_ nonterminal) SEED)))
\end{verbatim}

Next, define a Markov chain for each nonterminal: all make use of the same underlying substitution model
 (e.g. JC69 \cite{JukesCantor69}, HKY85 \cite{HasegawaEtAl85})
whose entries are stored as \verb|Q_a_b| for the transition rate between characters \verb|a| and \verb|b|.  
This ``underlying'' chain must be defined elsewhere - either in an included file (using the \verb|(&include)| directive), or directly in the grammar file.  
For instance, we could re-use the  JC69 chain, declaring rate parameters for later use:

\begin{verbatim}
(&foreach-token tok1
 (&foreach-token tok2
  (&if (&eq tok1 tok2) 
   ()  ;; If tok1==tok2, expand to an empty list (ignored by parser)
   (rate  (&cat Q_ tok1 _ tok2) u ))))
\end{verbatim}

Each \verb|nonterminal| has an associated substitution model which is \verb|Q_a_b| scaled by a different rate multiplier \verb|r_nonterminal|.  
Using an integer loop, we create a \verb|chain| for each nonterminal using the rate parameters we defined in the two previous code snippets:

\begin{verbatim}
(&foreach-integer nonterminal (1 numNonterms)
 (chain                   
  (terminal (&cat chain_ nonterminal))
  (&foreach-token tok1                  
   (&foreach-token tok2
    (&if (&eq tok1 tok2) 
     ()
     (mutate (from (tok1)) (to (tok2)) 
     (rate (&cat Q_ tok1 _ tok2) (&cat r_ nonterminal))))))))               
\end{verbatim}

Next, define the production rules which govern the nonterminal transitions.  For simplicity of  presentation (but not required), we assume here that 
transitions between nonterminals all occur with probability proportional to \verb|leaveProb|, and all self-transitions have probability \verb|stayProb|.   

The \verb|pgroup| declaration defines a probability distribution over a
 finite outcome space, with the parameters declared therein normalized
 to unity during parameter estimation. 
In this grammar we declare \verb|stayProb| and \verb|leaveProb|  within a \verb|pgroup| since they describe the two outcomes at
 each step of creating the alignment: staying at the current nonterminal or moving to a different one. 

\begin{verbatim}
(pgroup (stayProb 0.9) (leaveProb 0.1))
(&foreach-integer nonterm1 (1 numNonterms)
 ;; Each nonterminal has a transition from start
 (transform (from (start)) (to (nonterm1)) (prob (&/ 1 numNonterms))) 
 ;; Each nonterminal can transition to end - we assign this prob 1 
 ;; since the alignment length directs when this transition occurs                   
 (transform (from (nonterm1)) (to ()) (prob 1)) 
 (&foreach-integer nonterm2 (1 numNonterms)
  (&if (&eq nonterm1 nonterm2))
   ;; If nonterm1==nonterm2, this is a self-transition
   (transform (from (nonterm1)) (to (nonterm2)) (prob stayProb))            
   ;; Otherwise, this is an inter-nonterminal transition
   ;; with probability changeProb / (numNonterms - 1)
   (transform (from (nonterm1)) (to (nonterm2)) 
   (prob (&/ changeProb (&- numNonterms 1))))))
\end{verbatim}

Lastly, associate each nonterminal with its specially-designed Markov chain for emitted alignment columns:

\begin{verbatim}
(&foreach-integer nonterminal (1 numNonterms)       
 (transform (from (nonterminal))(to ((&cat chain_ nonterminal) (&cat nonterminal *))))
 (transform (from ((&cat nonterminal *))) (to (nonterminal))))
\end{verbatim}



\paragraph{Data-induced repetition}  
Models whose symmetric structure depends on the input data are less common in phylogenetic analysis, perhaps because normally their implementation  requires creating a new model for each new dataset to be analyzed.
  \xrate\ allows the user to create models based on different parts of the input data, namely the tree and the alignment, ``on the fly'' via its macro language.
  This is accomplished by making use of the tree iterators (e.g. \verb|&BRANCHES|, \verb|&NODES|, and  \verb|&LEAVES|) and alignment data (e.g. \verb|&COLUMNS|) to create nonterminals and/or terminal chains associated with these parts of the input data.  

In their program DLESS, Haussler and colleagues used such an approach in a tree-dependent model to detect lineage-specific selection. 
Their model used a phylo-HMM with different \hl{nonterminals} for each tree node, with the substitution rate below this node scaled to reflect gain or loss of functional elements \cite{SiepelEtAl2005}.  
We show a simplified form of their model as a schematic in \figref{dlessDiagram}, with blue colored branches representing a slowed evolutionary rate.  

Using \xrate's macros we can express this model in a compact way just as was done with the PhastCons model.  
Since both models use a set of nonterminals with their own scaled substitution models, we need
simply to replace the integer-based loop  \qquad \verb|(&foreach-integer nonterminal (1 numNonterms) expression)| \qquad
with the tree-based loop \qquad \verb|(&foreach-node state expression)| \qquad to create a nonterminal for each node in the tree.  
Then, define each node-specific chain as a {\em hybrid chain}, such that the chain associated with tree node $n$ has all the branches below node $n$ scaled to reflect heightened selective pressure.  
Hybrid chains, substitution processes which vary across the tree, are discussed briefly in the section on ``Recent enhancements to \xrate'', and the details 
of their specification is thoroughly covered in the \xrate\ format documentation, available here: \xrateformat\ . 
A minimal working form of the DLESS-style grammar included in \supptext{1}.


\subsubsection*{A repetitively-structured codon model specified using Scheme functions}
While \xrate's macro language is very flexible, there are some relatively common models that are difficult to express within the language's constraints.
  For example, a Nielsen-Yang codon matrix incorporating transition bias and selection has nearly 4,000 entries whose rates are determined  by the following criteria:

\[
Q^{NY}_{ij} = \left\{
\begin{array}{ll}
0 & \mbox{if $i$ and $j$ differ at more than one position}\\
\pi_j & \mbox{if $i$ and $j$ differ by a synonymous transversion}\\
\kappa\pi_j & \mbox{if $i$ and $j$ differ by a synonymous transition}\\
\omega\pi_j & \mbox{if $i$ and $j$ differ by a nonsynonymous transversion}\\
\omega\kappa\pi_j & \mbox{if $i$ and $j$ differ by a nonsynonymous transition}
\end{array}
\right.
\]

This sort of Markov chain is difficult to express in \xrate's macro language since its entries are determined by aspects of the codons (synonymous changes and transitions/transversions) which 
in turn depend on knowledge of the properties of nucleotides and codons that would have to be hard-coded directly into the loops and conditionals afforded by \xrate's macros. 
The conditions on the right side of the above equation are better framed as values returned from a function: given a pair of codons, the function
 returns the ``type'' of difference between them, which in turn determines the rate of substitution between the two codons.

\paragraph{Scheme extensions} It is this sort of situation which motivates extensions to \xrate\ that are more general-purpose than the simple macros described up to this point.
There are several valid choices for the programming language that can be used to implement such extensions.
For example, a chain such as $Q^{NY}$ can be generated fairly easily by way of a Perl or Python script tailored to generate \xrate\ grammar code.
While this is a convenient scripting mechanism for many users (and is perfectly possible with \xrate), 
it tends to lead to an awkward mix of code and embedded data (i.e. snippets of grammar-formatting text).
This obscures both the generating script and the final generated grammar file (the former due to the code/data mix, and the latter due to sheer size).  

Another choice of programming language for implementing \xrate\ extensions, which suffers slightly less
from these limitations, is Scheme.
As  \xrate's macro language is based on Lisp (the parent language to Scheme), the syntaxes are very similar, so the ``extension'' blends naturally with the surrounding \xrate\ grammar file.
Scheme is inherently functional and is also ``safe'' (in that it has garbage collection).
Lastly, data and code have equivalent formats in Scheme, enabling the sort of code/data mingling outlined above.  

To implement the $Q^{NY}$ chain in \xrate, we can use the \xrate\ Scheme standard library (found in \verb|dart/scheme/xrate-stdlib.scm|).
This standard library implements all the necessary functions to define the Nielsen-Yang model, with the genetic code implemented as a Scheme association list (facilitating easy substitution of alternate genetic codes, such as the mitochondrial code) as well as a wrapper function to initialize the entire model.

Without stepping through every detail of the Scheme implementation of the Nielsen-Yang model
in the \xrate\ standard library,
we will simply note that this implementation (the Nielsen-Yang model on a DNA alphabet)
is available via the following \xrate\ code
(the include path to {\tt dart/scheme} is searched by default
 by the Scheme function {\tt load-from-path}):

\begin{verbatim}
 (&scheme
  (load-from-path "xrate-stdlib.scm")
  xrate-dna-alphabet
  (xrate-NY-grammar))
\end{verbatim}

Note that {\tt xrate-dna-alphabet} is a simple variable,
but {\tt xrate-NY-grammar} is a function and is therefore wrapped in parentheses
(as per the syntax of calling a function in Scheme).
The reason that {\tt xrate-NY-grammar} is a function is so that the user can optionally
redefine the genetic code, which (as noted above) is stored as a Scheme association list,
in the variable {\tt codon-translation-table} (the standard library code can be examined for details).

%
%

\subsubsection*{A macro-heavy grammar for RNA structures in protein-coding exons}

As a final example of the possibilities that \xrate's new model-specification features enable, we present a new grammar for predicting RNA structures which  overlap protein-coding regions.  
XDecoder is based closely on the RNADecoder grammar first developed by Pederson and colleagues \cite{PedersenEtAl04}.
  This grammar is designed to detect phylogenetic evidence of conserved RNA structures, while also incorporating the evolutionary signals brought on by selection at the amino-acid level.
  In eukaryotes, RNA structure overlapping protein coding sequence is not yet well-known, but in viral genomes this is a common phenomenon due to constraints on genome size acting on many virus families.  
XDecoder is available as an \xrate\ grammar, linked here: \paperBiowikiPage

\paragraph{Motivation for implementation} Our endeavor to re-implement the RNADecoder grammar was based both on practical and methodological reasons.
  The original RNADecoder code is no longer maintained, but performs well on published viral datasets \cite{WattsEtAl2009}.
  Running RNADecoder on an alignment of full viral genomes is quite involved: the alignment must first be split up into appropriately-sized chunks (\~{}300 columns), converted to COL format \cite{COL},
and linked to a tree in a special XML file which directs the analysis.  
The grammar and its parameters, also stored in an XML format, are difficult to read and interpret.  
RNADecoder attains remarkably higher specificity in genome-wide scans as compared to protein-naive  prediction programs like PFOLD \cite{KnudsenHein2003}\ or MFOLD \cite{Zuker89b}.

\paragraph{Using XDecoder} We developed  our own \hl{variant} of the RNADecoder model as an \xrate\ grammar, called XDecoder.  This would have been a 
protracted task without \xrate's macro capabilities: the expanded grammar is nearly 4,000 lines of code.  
Using \xrate's macros, the main grammar (excluding the pre-estimated dinucleotide Markov chain)  is only \~{}100
 lines of macro code.
  Starting with an alignment of full-length {\em poliovirus} genomes, annotated with reading frames, an analysis can be run with a single simple command:

\noindent\verb| xrate -g XDecoder.eg -l 300 -wig polio.wig polio.stk > polio_annotated.stk| 

This runs \xrate\ with the XDecoder grammar on the Stockholm-format alignment \verb|polio.stk|,
allowing no more than 300 positions between paired columns, creating the wiggle file \verb|polio.wig|, annotating the original alignment with maximum likelihood secondary structure and rate class indicators,
and writing the annotated alignment to the the file \verb|polio_annotated.stk|.

Each analysis with RNADecoder requires an  XML file to coordinate the alignment and tree as well as direct parts of the analysis (training and annotation).
\xrate\ reads Stockholm format alignments which natively allows for alignment-tree association, enabling simple batch processing of many alignments.
The grammar can be run on arbitrarily long alignments, provided a suitable maximum pair length is specified via the \verb|-l N| argument. 
This prevents \xrate\ from considering any pairing whose columns are more than \verb|N| positions apart, effectively limiting both the memory usage and runtime.

Training the grammar's parameters, which may be necessary for running the grammar on significantly different datasets, is also accomplished with a single command:

\noindent\verb| xrate -g XDecoder.eg -l 300 -t XDecoder.trained.eg polio.stk|

The results of an analysis using XDecoder are shown in  \figref{grammar_compare}, together with gene and RNA structure annotations.  
Also shown are three related analyses (all done using \xrate\ grammars): PhastCons conservation, coding potential, and pairing probabilities computed using PFOLD.  
These three separate analyses reflect the signals that XDecoder must tease apart in order to reliably predict RNA structures.  
DNA-level conservation could be due to protein-coding constraints, regional rate variation, pressure to maintain a particular RNA structure, or a combination of all three.  
Using codon-position rate multipliers, multiple rate classes, and a secondary structure model, XDecoder unifies all of these signals in a single phylogenetic model, resulting in the highly-specific predictions shown at the top of \figref{grammar_compare}.





\subsection*{Recent enhancements to \xrate}
\subsubsection*{Lineage-specific models}
\seclabel{hybridchains}
All Markov chains in phylo-grammars describe the evolution of characters starting at the root and ending at the tips of the tree.  
In lineage-specific models, or {\em hybrid chains} in \xrate\ terminology, the requirement that all branches share the same substitution process is relaxed.  
Phylogenetic analysis is often used to detect a departure from a ``null model'' representing some typical evolutionary pattern.  
Standard applications of HMMs and SCFGs focus on modeling this departure on the alignment level, enabling different columns of the alignment to show different patterns of evolution. 
Using hybrid chains, users can explicitly model differences in evolution across  parts of the tree. By combining a hybrid chain with grammar nonterminals, this could be used to detect alignment regions \hl{(i.e. subsets of the set of all sites)} which display unusually high (or low) mutation rates in a particular part of the tree, 
such as in the DLESS model described in the section on ``Data-induced repetition''. 
The details of specifying such models are contained within the \xrate\ format documentation, at \xrateformat


\subsubsection*{Ancestral sequence reconstruction}
A phylo-grammar is a generative model: it generates a hidden parse tree, then further generates observed data conditional on that parse tree.
The observed data here is an alignment of sequences; the hidden parse tree describes which alignment columns are to be generated by the evolutionary models associated with which grammar nonterminals.
  Inference involves reversing the generative process: reconstructing the hidden parse structure and evolutionary trajectories that explain the alignment.

The original version of \xrate\ was focused on reconstructing the parse tree, for the purposes of annotating hidden structures such as gene boundaries or conserved regions.
A newly-implemented feature in \xrate\ allows an additional feature: reconstruction of ancestral sequences.
This functionality is already implicit in the phylogenetic model: no additional modification to the grammar is necessary to enable reconstruction.
The user can ask \xrate\ to return the most probable ancestral sequence at each internal node, or the entire posterior distribution over such sequences, via the \verb|-ar| and \verb|-arpp| command-line options.
\hl{Since} \xrate\ \hl{does marginal state reconstruction, the character with the highest
posterior probability returned by the} \verb|-arpp|  \hl{option will always correspond to 
the single character returned by the} \verb|-ar| \hl{option.}
Ancestral sequence reconstruction can be used to answer paleogenetic questions: what did the sequence of the  ancestor to all of clade $X$ look like?  
Similarly, evolutionary events such as particular substitutions or the gain or loss of function \hl{(also called trait evolution)} can be pinpointed to particular branches.  


\subsubsection*{Direct output of GFF and Wiggle annotations}
\xrate\ allows parse annotations to be written out directly in common bioinformatics file formats: GFF (a format for specifying co-ordinates of genomic features) \cite{GFF98}  
and WIG (a per-base format for quantitative data) \cite{WIG}.

This allows a direct link between \xrate\ and visualization tools such as JBrowse \cite{SkinnerEtAl2009}, GBrowse \cite{SteinEtAl2002}, the UCSC Genome Browser \cite{KentEtAl2002},and Galaxy \cite{pmid20738864}, allowing the results of different analyses to be displayed next to one another and/or processed in a unified framework. 

\paragraph{GFF: Discrete genomic features} GFF is a format oriented towards storing genomic features using 9 tab-delimited fields: each line represents a separate feature, with each field storing a particular aspect of the feature (e.g. identifier, start, end, etc).
With \xrate, a common application is using GFF to annotate an alignment with features corresponding to grammar nonterminals.
  For instance, using a gene-prediction grammar one could store the predicted start and end points of genes together with a confidence measure.
  Similarly, predicted RNA base pairs could be represented in GFF as one feature per pair, with start and end positions indicating the paired positions. 

\paragraph{WIG: Quantitative values for each column(s)} Wiggle format stores a  quantitative value for a single or group of positions. 
This can be especially useful to summarize a large number of possibilities as a single representative value.
  For instance, when predicting regions of structured RNA, \xrate\ may sum over many thousands of possible structures.
  We can summarize the model's results with the posterior probability that each column is involved in a base-pairing interaction.  





\subsection*{The Dart Scheme (Darts) interpreter}

Another way to use \xrate, instead of running it from the command line, is to call it from the Scheme interpreter (included in DART).  The compiled interpreter executable is named ``darts'' (for ``DART Scheme'').
This offers a simple yet powerful way to create parameter-fitting and genome annotation workflows.  For example, a user could train a grammar on a set of alignments, then use the resulting grammar to annotate a set of test alignments.

Darts, in common with the Scheme interpreter used in \xrate\ grammars,
is implemented using Guile (GNU's Ubiquitous Intelligent Language for Extension:
{\tt http://www.gnu.org/software/guile/guile.html}).
Certain commonly-encountered bioinformatics objects,
serializable via standard file formats and implemented as C++ classes within \xrate,
are exposed using Guile's ``small object'' (smob) mechanism.
Currently, these types include Newick-format trees and Stockholm-format alignments.
API calls are provided to construct these ``smobs'' by parsing strings (or files) in the appropriate format.
The smobs may then be passed directly as parameters to \xrate\ API calls, or may be ``unpacked''
into Scheme data structures for individual element access.
Guile encourages sparing use of smobs; consequently,
smobs are used within Darts exclusively to implement bioinformatic objects that already have a broadly-used file format
(Stockholm alignments and Newick trees).
In contrast, formats that are newly-introduced by \xrate\ (grammars, alphabets and so forth)
are all based on S-expressions, and so may be represented directly as native Scheme data structures.

The functions listed in \secref{scheme_functions} provide an interface between Scheme and \xrate.  Together with the functions in the \xrate-scheme standard library and Scheme's native functional scripting abilities,  a broad array of models and/or workflows are possible.  For instance, one could estimate several sets of parameters for Nielsen-Yang models using groups of alignments, and then embed each one in a PhastCons-style phylo-HMM, finally using this model to annotate a set of alignments.  While this and other workflows could be accomplished in an external framework (e.g. Make, Galaxy \cite{pmid20738864}), Darts provides an alternate way to script \xrate\ tasks using the same language that is used to construct the grammars.

\section*{Materials and Methods}
\supptext{1} contains example grammars referred to in the text, as well as small  and large test Stockholm alignments.  
The alignment of {\em poliovirus} genomes along with the grammars used to produce \figref{grammar_compare} are also included along with a Makefile indicating how the data was analyzed.
Typing {\tt make help} in the directory containing the Makefile will display the demonstrations available to users. 

\section*{Acknowledgments}

\bibliography{allbib}

\section*{Figure Legends}

\begin{center}
\begin{figure}[h!]
\includegraphics[width=1\textwidth]{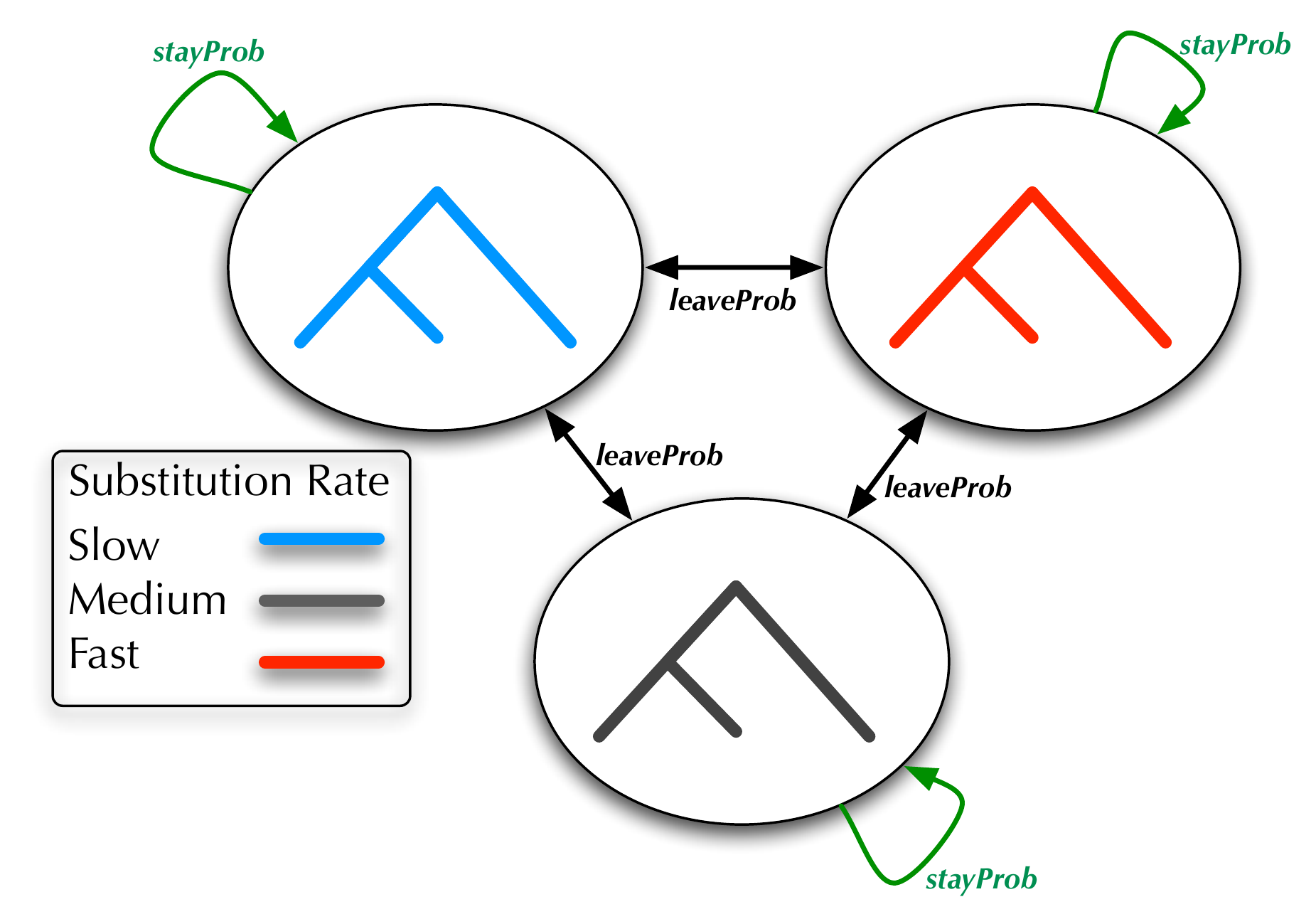}
\caption{
{\bf The model used by PhastCons,  a 3-nonterminal HMM with rate multipliers, is compactly expressed by \xrate's macro language. }
Different nonterminal have different evolutionary rates, but they all share the same underlying substitution model.  
Transition probabilities are shared: a transition between nonterminals happens with probability {\em leaveProb}, and self-transitions happen with probability {\em stayProb}.  
This model (with any number of nonterminals) can be expressed in \xrate's macro language in approximately 20 lines of code.
}
\figlabel{phastconsDiagram}
\end{figure}
\end{center}

\begin{center}
\begin{figure}[h!]
\includegraphics[width=1\textwidth]{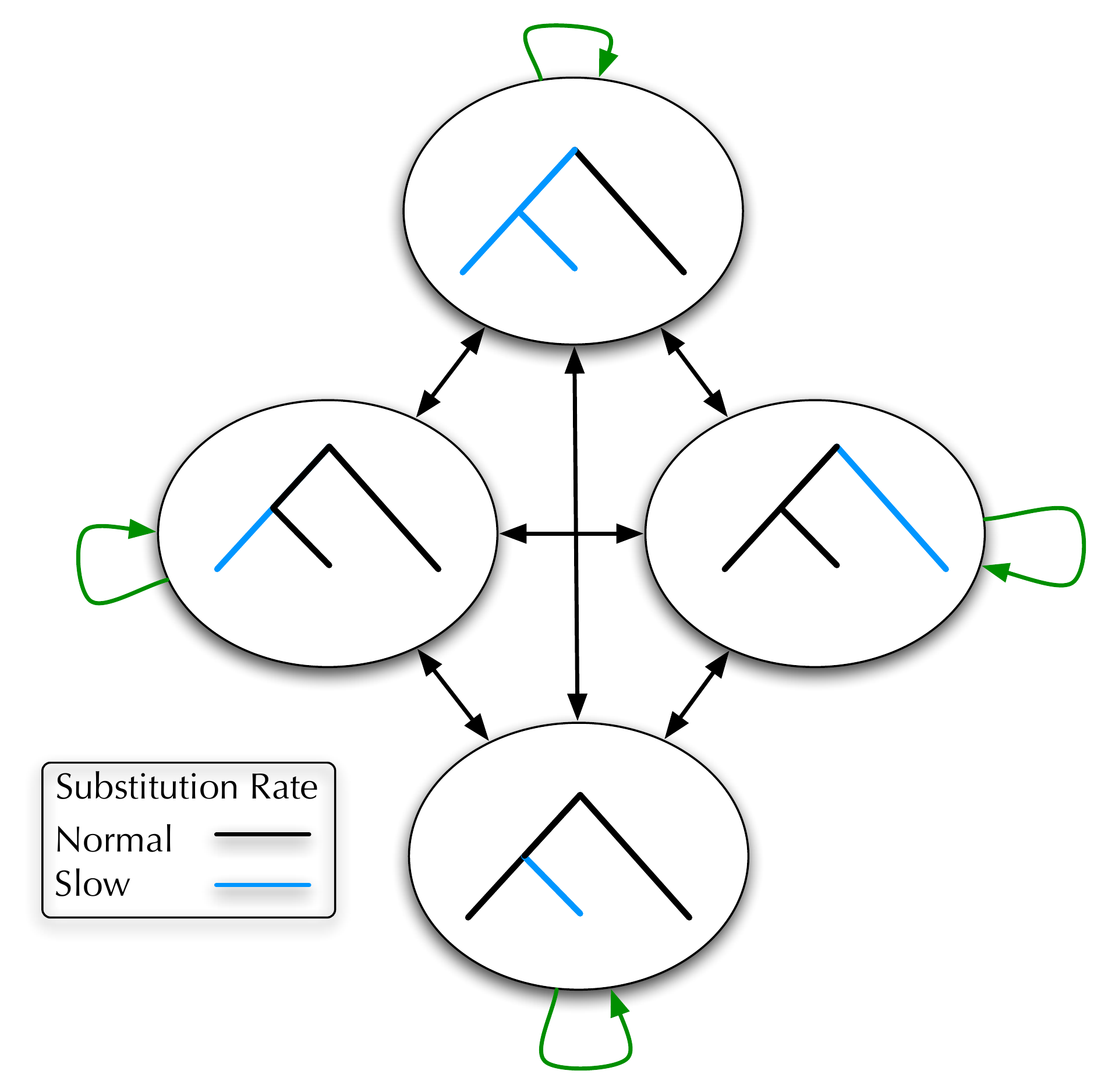}
\caption{
{\bf A schematic of a DLESS-style phylo-HMM: each node of the tree has its own nonterminal, such that the node-rooted subtree evolves at a slower rate than the rest of the tree.}  Inferring the pattern of hidden nonterminals generating an alignment allows for detecting regions of lineage-specific selection.  Expressing this model compactly in \xrate's macro language allows it to be used with any input tree without having to write data-specific code or use external model-generating scripts. 
}
\figlabel{dlessDiagram}
\end{figure}
\end{center}

\begin{center}
\begin{figure}[h!]
\includegraphics[width=1\textwidth]{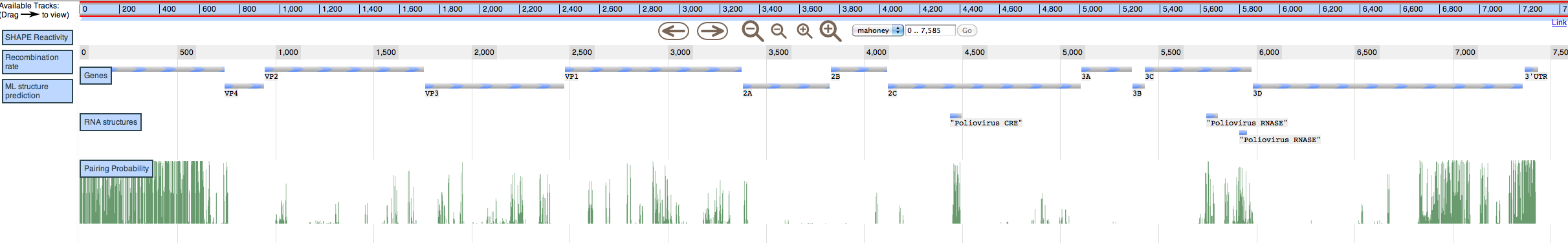}
\caption{
{\bf Data from several \xrate\ analyses, shown alongside  genes {\bf (A)} and known RNA structures {\bf (B)} in {\em poliovirus}.}
XDecoder  {\bf (C)} recovers all known structures with high posterior probability and  predicts a  promising target for experimental probing (region 6800-7100).  
XDecoder was run on an alignment of 27 {\em poliovirus} sequences with the results visualized as a track in JBrowse \cite{SkinnerEtAl2009} via a wiggle file. 
Alongside XDecoder probabilities are the three signals which XDecoder aims to disentangle: {\bf (D)}  conservation, {\bf (E)} coding potential, and {\bf (F)} RNA structure.  
Paradoxically, the CRE and RNase-L inhibition elements show both conservation and coding sequence preservation, whereas PFOLD's predictions show only a slight increase in probability density around the known structures.  XDecoder is the only grammar which returns predictions of reasonable specificity.  The full JBrowse instance is included  as \supptext{2}. 
}
\figlabel{grammar_compare}
\end{figure}
\end{center}

\section*{Tables}
\appendix

\section{Glossary of XRate model terminology}
\seclabel{glossary}
Within the glossary descriptions, {\em italicized phrases} refer to other glossary terms.
\begin{description}
\item[Alignment:] See {\em multiple sequence alignment}.
\item[Alphabet:] The set of single-character tokens (symbols) from which sequences are constituted.
The alphabet is defined in the grammar file; only one alphabet may be defined per grammar file.
(Usually the alphabet is DNA, RNA or protein. Sometimes the alphabet is extended to include an explicit gap character.)
An alphabet may optionally include a {\em complement} mapping,
as well as specification of degenerate (ambiguous) characters.
\item[Ancestral reconstruction:] The use of \xrate\ to reconstruct the sequences at ancestral nodes
of a {\em phylogenetic tree}, given a {\em grammar}, a {\em multiple sequence alignment}
and a {\em parse tree}.
This occurs after {\em tree estimation}, {\em training} and {\em annotation}.
\item[Annotation:] The use of \xrate\ to apply a {\em grammar} to a {\em multiple sequence alignment}
and {\em phylogenetic tree},
so as to impute the optimal {\em parse tree}
and mark up the alignment with the co-ordinates of selected features (associated with particular {\em nonterminals}
in the parse tree), or generate other annotation including GFF and WIGgle files.
This occurs after {\em tree estimation} and {\em training}, but prior to {\em ancestral reconstruction}.
Can also refer to a specific part of a {\em transformation rule} that generates annotations.
\item[Bifurcation:] A {\em transformation rule} that generates two {\em nonterminals}.
Bifurcation rules have the form \\
\centerline{\tt (transform (from (A)) (to (B C)))}
where {\tt A}, {\tt B} and {\tt C} are nonterminals.
\item[Chain:] A substitution rate matrix (the name comes from ``continuous-time Markov chain'').
The states of the substitution process are $N$-mers augmented with an optional hidden variable.
That is, the state space of a chain consists of {\em state-tuples} of the form $(s_1,s_2,\ldots,s_N,h)$
with $N \geq 1$,
where $s_1$ through $s_N$ represent {\em alphabet} symbols
(which will be observed in the final {\em multiple sequence alignment})
and $h$ is an optional {\em hidden state} which can take on
a finite set of single-character values specific to this chain.
Each of the $N$ alphabet symbols, $s_1$ through $s_N$, is associated with a unique {\em pseudoterminal}.
Examples of valid chain state spaces include the set of all nucleotides; the set of all codons;
and the set of all tuples $(A,H)$ where $A$ is an amino acid and $H \in \{ F, S \}$ is a hidden binary variable
taking values $F$ (for fast) or $S$ (for slow).
\item[Complement:] An order-2 permutation on the {\em tokens} of an {\em alphabet}.
(Typically only used for DNA or RNA alphabets.)
\item[Emission:] A {\em transformation rule} that generates some {\em pseudoterminals}
(and thus, some alignment columns);
or the set of pseudoterminals (or alignment columns) generated by such a rule.
In \xrate, emission rules have the form $A \to x_1 \ldots x_L\ {A*} \ x_{L+1} \ldots x_{L+R}$
where $A,A*$ are paired {\em nonterminals} (whose names differ only by the final asterisk)
and $x_1 \ldots x_{L+R}$ are pseudoterminals.
($A*$ is referred to as the {\em post-emit nonterminal}.)
Any numbers $L,R$ of pseudoterminals can appear to the left and right of the $A*$,
as long as $L+R>0$.
If $L=0$ and $R>0$, the rule is a {\em right-emission};
if $L>0$ and $R=0$, the rule is a {\em left-emission}.
The pseudoterminals $x_1 \ldots x_{L+R}$ must comprise (any permutation of)
the full set of pseudoterminals for a given {\em substitution chain}.
Each pseudoterminal may optionally be prefixed with a tilde character ({\tt \~{}}) to indicate that it
should be {\em complemented} in the final alignment (used to generate reverse strands in double-stranded models).
For example, if {\tt C1}, {\tt C2} and {\tt C3} are the three pseudoterminals of a codon chain,
{\tt A} is an emission nonterminal and {\tt A*} is the corresponding post-emission nonterminal,
valid emission rules could include \\
\centerline{\tt (transform (from (A)) (to (C1 C2 C3 A*)) (prob (...)))}
and \\
\centerline{\tt (transform (from (A)) (to (\~{}C3 A* \~{}C2 \~{}C1)) (prob (...)))}
\item[Grammar:] The contents of a grammar file:
{\em chains}, {\em nonterminals}, {\em transformation rules}, and {\em alphabet}.
(The alphabet is specified in a separate part of the file
from the rest of the grammar, and so is sometimes omitted from this definition.)
\item[Grammar symbol:] A symbol that is either a {\em nonterminal} or a {\em pseudoterminal}.
\item[HMM:] Hidden Markov Model. An {\em SCFG} that is also a {\em regular grammar}. See also {\em phylo-HMM}.
\item[Hidden state:] In the context of \xrate,
this term is ambiguous (see {\em state}).
In this article, it is used mostly to refer to the final element of a {\em state-tuple} in a {\em chain}.
However, in the context of {\em HMM} theory, it refers to what we call a {\em nonterminal}.
\item[Hybrid chain:] A mapping from tree branches to substitution rate matrices ({\em chains}) where the
instantaneous rate matrix may vary from one branch to another.
This may be used to implement lineage-dependent selection,
or other models which are heterogeneous with respect to the tree.
\item[Initial distribution:] The initial probability distribution over states in a substitution {\em chain}.
\item[Left-emission:] See {\em emission}.
\item[Left-regular:] A {\em grammar} is left-regular if it contains no {\em bifurcations}
and its {\em emissions} are all {\em left-emissions}.
\item[Macro:] A construct that is expanded by the \xrate\ grammar preprocessor and may be used to
implement redundant or repetitive grammar models; e.g. {\em grammars} with a large number
of similar {\em transformation rules} sharing the same probability parameter,
or substitution {\em chains} whose {\em mutation rules} all share the same rate parameter.
\item[Multiple sequence alignment:] The raw data on which \xrate\ operates,
and which constitutes its input and output.
\xrate\ cannot align sequences, but assumes that they have been pre-aligned using an external alignment program.
Alignments must be converted to Stockholm format \cite{StockholmFormat} before supplying them to \xrate.
The alignment may include a {\em phylogenetic tree} (using the Stockholm syntax for specifying this);
if no tree is provided, \xrate's {\em tree estimation} routines can be used to find one.
\item[Mutation rule:] A single element in the rate matrix of a substitution {\em chain}.
\item[Nonterminal:] A {\em grammar symbol} that may be transformed, by application of {\em transformation rules},
into other nonterminals or pseudoterminals.
In \xrate, a nonterminal must be exclusively associated with
(that is, appear on the left-hand side of)
either {\em emission} rules, {\em transition} rules or {\em bifurcation} rules.
\item[Parameter:] A named parameter in a grammar. May be a {\em probability parameter} or a {\em rate parameter}.
\item[Parametric model:] A {\em grammar} whose {\em transformation rules} or {\em mutation rules} (or both)
are specified as functions of the grammar's {\em parameters}, rather than as direct numerical values.
\item[Parse tree:] A tree structure corresponding to the derivation of a multiple sequence alignment
from a {\em grammar}. Each tree node is labeled with a {\em grammar symbol}: the root node is labeled with
the {\em start nonterminal}, internal nodes are labeled with {\em nonterminals}, and the leaves are
labeled with {\em pseudoterminals}.
Not to be confused with a {\em phylogenetic tree}.
\item[PGroup:] A set of {\em probability parameters} collectively representing a probability distribution
over a finite set of events. Following training, probability parameters constituting a PGroup will be normalized
to sum to 1.
\item[Phylogenetic tree:] The evolutionary tree describing the relationship between sequences in a multiple alignment.
\xrate\ uses the Stockholm format for alignments, which allows the tree to be included as an annotation of the
alignment.
If no tree is provided, \xrate's {\em tree estimation} routines can be used to find one.
\item[Phylo-grammar:] See {\em phylo-SCFG}.
\item[Phylo-HMM:] A {\em phylo-SCFG} that uses a {\em regular grammar}.
A phylo-HMM is an HMM whose {\em emissions} generate alignment columns
by evolving {\em substitution chains} on a phylogenetic tree.
\item[Phylo-SCFG:] A phylogenetic {\em SCFG}: a member of the general class of {\em grammars} implemented by \xrate.
A phylo-SCFG is an SCFG whose {\em emissions} generate alignment columns
by evolving {\em substitution chains} on a phylogenetic tree.
\item[Post-emit nonterminal:] See {\em emission}.
\item[Production rule:] See {\em transformation rule}.
\item[Probability parameter:] A dimensionless {\em parameter} that generally takes a value between 0 and 1,
and so can occur in the probability part of a {\em transformation rule}
(or as a multiplying factor in the rate part of a {\em mutation rule}).
Probability parameters are declared in {\em PGroups}.
\item[Pseudocounts:] A set of nonnegative counts that specifies a Dirichlet prior distribution over a {\em PGroup}.
\item[Pseudoterminal:] A {\em grammar symbol} that is generated via an {\em emission}
and cannot be further modified by subsequent {\em transformation rules}.
In a {\em parse tree}, a pseudoterminal serves as a placeholder for an alignment column.
Pseudoterminals occur in groups associated with a particular {\em substitution chain}.
In the generative interpretation of the model,
alignment columns are generated using the {\em initial distribution} and {\em mutation rules} of the chain,
applied on the phylogenetic tree associated with the alignment.
\item[Rate parameter:] A nonnegative parameter that has units of ``inverse time'' (i.e. rate),
and so can occur in the rate part of a {\em mutation rule}.
Rate parameters can be declared individually.
\item[Regular grammar:] A {\em grammar} is regular if it is either {\em left-regular} or {\em right-regular};
that is, it contains no {\em bifurcations} and its {\em emissions} are all either
{\em left-emissions} or {\em right-emissions}.
A regular grammar is equivalent to an {\em HMM}.
\item[Right-emission:] See {\em emission}.
\item[Right-regular:] A {\em grammar} is right-regular if it contains no {\em bifurcations}
and its {\em emissions} are all {\em right-emissions}.
\item[SCFG:] Stochastic Context-Free Grammar. See also {\em phylo-SCFG}.
\item[Start nonterminal:] The first {\em nonterminal} declared or used in a grammar.
In the generative interpretation of the model, this is the initial {\em grammar symbol} to which transformation rules
are applied. It is also the label of the root node in the {\em parse tree}.
\item[State:] In the context of a {\em phylo-grammar},
this term is ambiguous: it can refer either to a {\em state-tuple} in a {\em chain},
or (for phylo-HMMs) a {\em nonterminal} in a {\em grammar}.
For the most part in this paper, and exclusively in this glossary, we use it in the former sense.
\item[State space:] The set of possible {\em state-tuples} in a {\em chain}.
\item[State-tuple:] A tuple of the form $(s_1,s_2,\ldots,s_N,h)$
representing a single state in a {\em chain},
where $s_1$ through $s_N$ represent {\em alphabet} symbols
and $h$ is an optional {\em hidden state}.
\item[Substitution chain:] A continuous-time finite-state Markov chain over {\em state-tuples}. See {\em chain}.
\item[Substitution model:] See {\em substitution chain}.
\item[Terminal:] See {\em token}.
\item[Token:] An {\em alphabet} symbol. (Also called a {\em terminal}.)
\item[Training:] The use of \xrate\ to estimate a grammar's {\em parameters}, {\em mutation rule} rates
and {\em transformation rule} probabilities, given a (set of) multiple alignments.
This occurs after {\em tree estimation} and prior to {\em annotation} or {\em ancestral reconstruction}.
\item[Transformation rule:] A probabilistic rule that describes the transformation of a {\em nonterminal} symbol
into a sequence of zero or more {\em grammar symbols}. (Also called a {\em production rule}.)
A transformation rule may be an {\em emission}, a {\em transition} or a {\em bifurcation}.
\item[Transition:] A {\em transformation rule} that generates exactly one nonterminal (and no pseudoterminals).
Transition rules have the form \\
\centerline{\tt (transform (from (A)) (to (B)) (prob (...)))}
where {\tt A} and {\tt B} are nonterminals.
\item[Tree:] In the context of a {\em phylo-grammar}, this term is ambiguous: it can mean a
{\em parse tree} (which explains the ``horizontal'', i.e. spatial, structure of an alignment)
or a {\em phylogenetic tree} (which explains the ``vertical'', i.e. temporal, structure).
\item[Tree estimation:] The use of \xrate\ to estimate a {\em phylogenetic tree}
for a {\em multiple sequence alignment}, given a {\em grammar}.
This occurs prior to {\em training}, {\em annotation} or {\em ancestral reconstruction}.
\end{description}

\section{Tables of Scheme functions in Darts}
\seclabel{scheme_functions}
The following list of Scheme functions, natively implemented within selected DART programs
(including XRate) when compiled with GNU Guile, is only complete to the date of publication.
A more up-to-date list may be found at \url{http://biowiki.org/DartSchemeFunctions}
\subsection{Functions for working with trees}
\begin{tabular}{lp{5cm}}
Scheme function & Effect  \\
\hline \\
(newick-from-string x) & Create a tree-smob from a Newick-format string x \\
(newick-from-file x) & Create a tree-smob from a file x \\
(newick-from-stockholm x) & Create a tree-smob from the tree encoded within alignment-smob x \\
(newick-to-file x y) & Write tree-smob x to file y in Newick format \\
(newick-ancestor-list x) & List of all ancestors in the tree-smob x\\
(newick-leaf-list x) & List of all leaves in the tree-smob x \\
(newick-branch-list x) & List of all branches in the tree-smob x \\
(newick-unpack x) & Converts a tree-smob x into a Scheme data structure \\
\end{tabular}

\subsection{Functions for working with alignments}
\begin{tabular}{lp{5cm}}
Scheme function & Effect  \\
\hline \\
(stockholm-from-string x) & Create an alignment-smob from a Stockholm-format string x  \\
(stockholm-from-file x) & Create an alignment-smob from a Stockholm-format file x \\
(stockholm-to-file x y) & Write alignment-smob x to Stockholm-format alignment file y \\
(stockholm-column-count x) & Return the number of columns in alignment-smob x \\
(stockholm-unpack x) & Converts an alignment-smob x into a Scheme data structure \\
\end{tabular}

\subsection{Functions for working with grammars}
\begin{tabular}{lp{5cm}}
Scheme function & Effect  \\
\hline \\
(xrate-validate-grammar x) & Validate the syntax of \xrate\ grammar x \\
(xrate-validate-grammar-with-alignment x y) & Validate the syntax of \xrate\ grammar x, using alignment-smob y to expand macro constructs \\
(xrate-estimate-tree x y) & Use \xrate\ grammar y to estimate a tree for alignment-smob x \\
(xrate-annotate-alignment x y) & Use \xrate\ grammar y to annotate alignments-smob x \\
(xrate-train-grammar x y) & Train \xrate\ grammar y on the list of alignment-smobs y \\
\end{tabular}

\subsection{Miscellaneous functions}
\begin{tabular}{lp{5cm}}
Scheme function & Effect  \\
\hline \\
(dart-log x) & Logging directive; equivalent to ``{\tt -log x}'' at the command line \\
(discrete-gamma-medians alpha beta K) & Returns the median rates of K equal-probability bins of the gamma distribution \cite{Yang94}  \\
(discrete-gamma-means alpha beta K) & Returns the mean rates of K equal-probability bins of the gamma distribution \cite{Yang94}  \\
(ln-gamma k) & Calculates the gamma function, $\Gamma(k) = \int_0^\infty e^{-x} x^{k-1} dx$ \\
(gamma-density x alpha beta) & Calculates the gamma probability density, $\beta^\alpha \frac{1}{\Gamma(\alpha)} x^{\alpha-1} e^{-\beta x}$ \\
(incomplete-gamma x alpha beta) & Calculates the incomplete gamma function, i.e. the integral of the gamma density up to x \\
(incomplete-gamma-inverse p alpha beta) & Calculates the inverse of the incomplete gamma function \\
\end{tabular}

\end{document}